# Exotic magnetic anisotropy near digitized dimensional Mott boundary

*Seung Gyo Jeong, Jihyun Kim, Taewon Min, Sehwan Song, Jin Young Oh, Woo-suk Noh, Sungkyun Park, Tuson Park, Jong Mok Ok, Jaekwang Lee\*, and Woo Seok Choi\**

S. G. Jeong, J. Kim, J. Y. Oh, T. Park, W. S. Choi
Department of Physics, Sungkyunkwan University, Suwon 16419, Korea
E-mail: choiws@skku.edu

T. Min, S. Song, S. Park, J. M. Ok, J. Lee
Department of Physics, Pusan National University, Busan 46241, Korea
E-mail: jaekwangl@pusan.ac.kr

W.-s. Noh
MPPC-CPM, Max Planck POSTECH/Korea Research Initiative, Pohang 37673, Korea

The magnetic anisotropy of low-dimensional Mott systems exhibits unexpected magnetotransport behavior useful for spin-based quantum electronics. Yet, the anisotropy of natural materials is inherently determined by the crystal structure, highly limiting its engineering. We demonstrate the magnetic anisotropy modulation near a digitized dimensional Mott boundary in artificial superlattices composed of a correlated magnetic monolayer SrRuO$_3$ and nonmagnetic SrTiO$_3$. The magnetic anisotropy is initially engineered by modulating the interlayer coupling strength between the magnetic monolayers. Interestingly, when the interlayer coupling strength is maximized, a nearly degenerate state is realized, in which the anisotropic magnetotransport is strongly influenced by both the thermal and magnetic energy scales. Our results offer a new digitized control for magnetic anisotropy in low-dimensional Mott systems, inspiring promising integration of Mottronics and spintronics.

# 1. Introduction

Low-dimensional Mott phases give rise to spin-correlated quantum phenomena such as current-induced magnetism, quantum criticality, colossal magnetoresistance (MR), quantum oscillation, Higgs mode, and quantum spin liquid phase with gapless Majorana fermions.[1-6] Whereas the conventional Mott transition is determined by the competition between the on-site Coulomb potential ($U$), electronic bandwidth ($W$), exchange coupling ($J$), and spin-orbit coupling ($\xi$), the low-dimensional Mott phase is further characterized by the quantum confinement effect, quantum fluctuation, Rashba effect, and anisotropic orbital hybridization.[7-11] Importantly, the Mott metal-insulator transition (MIT) is frequently coupled to the spin-state transition established by the Kanamori Hamiltonian,[12] which is strongly influenced by dimensionality.

Long-range spin order can also be realized in low-dimensional Mott systems, in which magnetic anisotropy plays a crucial role in realizing novel electromagnetic phases and resultant MR. Initially, magnetic anisotropy is necessary to break the degeneracy of spin-wave states and stabilize the long-range spin order in two-dimensional (2D) magnetic crystals.[13] This anisotropic 2D spin order inherently promotes the anisotropic magnetotransport. For example, strong $\xi$ in monolayer $CrI_3$ and $VI_3$ introduces magnetic anisotropy and the resultant ferromagnetic ordering within the 2D Mott insulating phases,[14] frequently accompanying a large anisotropic MR. Layered perovskite $Ca_3Ru_2O_7$ exhibits an A-type antiferromagnetic insulating phase and shows Mott MIT depending on the magnetic anisotropy modulated by the direction of the external magnetic-($H$-)field.[2] $Sr_2IrO_4$ is also an antiferromagnetic Mott insulator, where the interlayer interaction strength is five orders of magnitude smaller than that of intralayer interactions.[15] This allows the $H$- or pressure-induced antiferromagnetic to weak ferromagnetic transitions accompanied by an electronic phase transition. Such magnetic anisotropy modulations may assist developments of future spintronic devices, including magnetic memories, spin-based logic devices, magnetic sensors, magnetic resonance imaging, spin-torque oscillators, spin valves, and quantum computation. However, natural layered materials have limitations in actively modulating magnetic anisotropy because the anisotropy is intrinsically predetermined by the low-dimensional crystal structure.

Atomic-scale precision epitaxy of perovskite oxide superlattices offers opportunities for controlling magnetic anisotropy within a synthetic low-dimensional Mott system. The vast material possibilities and delicate thickness control in superlattices offer facile control knobs of correlated functionalities, including dimensionality-induced MIT, magnetic phase transition, exchange bias, and magnetic anisotropy engineering.[16-20] One way of measuring the anisotropy in low-dimensional synthetic crystals would be to assess the ratio between the intralayer and interlayer interactions. For a superlattice composed of monolayers of ferromagnetic metal and sufficiently thick layers of nonmagnetic insulator, interlayer coupling between the magnetic monolayers is expected to disappear, leading to 2D confinement (**Figure 1**a). When the thickness of the nonmagnetic insulator layers decreases, the interlayer coupling between the magnetic monolayers begins to increase, whereas the intralayer coupling remains essentially unchanged. Even with the same $U$ value for the ferromagnetic layer, the electronic structure of the system approaches that of a three-dimensional (3D) system with a broader $W$, as the interlayer coupling strength is enhanced (Figure 1b). Therefore, effective dimensionality control of anisotropy would realize an MIT strongly coupled to the magnetic state in synthetic 2D Mott systems. SrRuO$_3$ (SRO) films show Stoner ferromagnetism with perpendicular magnetic anisotropy in general, and further exhibit low-dimensional Mott phases with magnetic ordering.[1, 16, 21] Hence, if accessible, dimensionality control of the monolayer SRO near the Mott phase boundary provides us with the opportunity of accessing exotic magnetic ground states closely linked to the MR.

In this study, we discovered an intriguing anisotropy modulation of MR in a synthetic low-dimensional Mott system. We fabricated superlattices composed of monolayer SRO and a nonmagnetic insulator SrTiO$_3$ (STO). The interlayer coupling strength was digitally controlled by the atomic-scale precision control of $y$ within the [1|$y$] superlattice (one- and $y$-unit cells of the SRO and STO layers, respectively) (see Methods and Figure S1 in the Supporting Information). This led to an effective engineering in the dimensionality which is closely associated with the Mott transition in SRO. Hence, a digitized dimensional Mott boundary is realized, at which we examined the interlayer coupling-induced Mott transition coupled to the magnetic transition, as schematically shown in Figure 1a, b. We further observed an effective anisotropy engineering of MR for 2D SRO crystals near the dimensional Mott boundary, depending on the $H$-field, temperature ($T$), and interlayer coupling.

## 2. Results and Discussions

The dimensional Mott MIT with coupled magnetic transition in 2D SRO crystals is demonstrated by density functional theory (DFT) calculations, as shown in Figure 1c. The [1|6] ($y = 6$) superlattice with sufficiently thick STO layers shows an antiferromagnetic Mott insulating phase with a small $W$ as the ground state of 2D SRO, consistent with previous studies on atomically thin SRO layers.[16, 21-25] With decreasing $y$ ($y = 4$ and 2), $W$ increases and the broader bands near the Fermi level merge, eventually forming a half-metallic state for the [1|1] ($y = 1$) superlattice, similar to the 3D bulk SRO. The enhanced interlayer coupling between SRO monolayers also induces a magnetic phase transition from an antiferromagnetic phase for $y = 6$ to a ferromagnetic phase for $y = 1$ superlattices, which simultaneously occurs with Mott MIT. The ferromagnetic (antiferromagnetic) phase of SRO strongly favors metallic (insulating) states from our DFT calculations, suggesting a close correlation between the spin and charge degrees of freedom in the low-dimensional SRO system.[16, 22] Since the DFT calculations of the [1|$y$] superlattices have used identical crystalline symmetry (tetragonal for the thin SRO and cubic for the STO layers, as observed from the experimental structural characterization (Figure S2, Supporting Information)) and $U$ values of 2.6 eV, the magnetically coupled Mott transition originates primarily from the interlayer coupling strength modulation between the SRO monolayers.

Figure 1d-f shows experimental confirmation of the dimensional Mott MIT coupled to the magnetic transition. $T$-dependent resistivity ($\rho$ ($T$)) curves confirm the $y$-dependent digitized dimensional MIT of the [1|$y$] superlattices (Figure 1d). In particular, in the low-$T$ region, the electronic ground states of the $y = 4$ and 6 superlattices exhibit $\rho$ ($T$) above the Mott-Ioffe-Regel limit,[25] indicating the Mott insulating states. On the other hand, the $y = 1$ and 2 superlattices show semiconducting behavior with the $\rho$ ($T$) values within the Mott-Ioffe-Regal limit at the low-$T$ region. As shown in Figure 1e, $T$-dependent magnetization ($M$ ($T$)) curves along the perpendicular direction exhibit the clear appearance of ferromagnetic transition temperature for the $y = 1$ and 2 superlattices at 100 and 80 K, respectively. $H$-field-dependent $M$ ($M$ ($H$)) curves at 5 K further support the interlayer coupling-induced long-range spin order (Figure 1f). While the saturation $M$ values at 5 T are similar for all [1|$y$] superlattices (1.7-1.9 $\mu_B$/Ru) and bulk and thin film SRO samples,[25] remnant $M$ and coercive $H$-field along the perpendicular direction enhance considerably with decreasing $y$ below 4, (also see the insets of Figure 1f for the opening of the ferromagnetic hysteresis loop, Figure S3 and S4 for the ferromagnetic magnetotransport behavior, and Figure S5 for finite X-ray magnetic circular

dichroism (XMCD) signal at Ru-$L_{2,3}$ edge in the Supporting Information). Note that qualitatively similar $M$ ($H$) curves are shown for the $H$-field along the horizontal direction but with highly suppressed coercivity, highlighting the anisotropic nature. In addition, MR ($H$) curves of $y$ = 1 superlattices at 5 K confirm the anisotropic ferromagnetic ordering with finite and nearly zero coercive fields (characterized by $H$-field at maximum MR ($H_{peak}$)) for the perpendicular (MR$_p$) and horizontal (MR$_h$) $H$-fields, respectively (Figure S4, Supporting Information). The interlayer coupling-induced ferromagnetic metallic phase exhibits a unique low-dimensional magnetotransport behavior. When the interlayer coupling strength within the SRO monolayer superlattices ($y$ = 1) is maximized, an unexpected anisotropy in MR emerges, which is highly susceptible to both the thermal and magnetic energy scales.

**Figure 2** shows anisotropic magnetotransport at the selected $H$-field and $T$ in the 2D SRO crystal. First, we focus on the $\gamma$-dependent angular magnetoresistance (AMR) at $\alpha$ = 90° and $\beta$ = 0° (three azimuthal angles of the applied $H$-field, $\alpha$, $\beta$, and $\gamma$, are defined in the schematic diagram in the second row of Figure 2.) for the $y$ = 1 superlattice. At a low $H$-field (1 T) (Figure 2a), the minimum AMR (AMR ($\gamma_{min}$), where $\gamma_{min}$ is $\gamma$ angle of minimum AMR) develops at $\gamma$ = 90° as $T$ decreases below 20 K, with peak structures at $\gamma$ = 0° and 180°. Surprisingly, as the $H$-field (9 T) increases at low $T$ (5 K) (Figure 2b), peaks start to emerge at $\gamma$ = 90°, shifting AMR ($\gamma_{min}$) towards 30° and 150° approximately, while maintaining the peak structures at $\gamma$ = 0° and 180°. The unexpected $H$-field-dependent AMR can be decoupled by considering distinctive anisotropic spin ordering (see Text S1 in the Supporting Information). Conventionally, AMR ($\gamma_{min}$) suggests spin ordering along the $H$-field, which is aligned toward $\gamma_{min}$. Hence, the perpendicular (horizontal) spin ordering induces $\gamma_{min}$ = 0° (90°). As shown in Figure 2c, the combination of perpendicular (light blue line) and horizontal (purple line) spin ordering reproduced the experimentally observed AMR. In particular, the perpendicular and horizontal AMR can be phenomenologically described as $p(|1 - \cos(\gamma)|)$ and $-h|\sin(\gamma)|$, respectively, where $p$ and $h$ are scale constants. The $\gamma$-dependent AMR of perpendicular spin ordering ($\gamma_{min}$ = 0°), which is frequently observed in conventional 3D SRO films,[19, 27, 28] follows $|1 - \cos(\gamma)|$ (Figure S6, Supporting Information), justifying the use of the absolute sinusoidal function for the AMR. As shown for the case of AMR at 9 T and 5 K, the combination result ($p(|1 - \cos(\gamma)| - h|\sin(\gamma)|)$) is in excellent agreement with the experiment, except for the hysteresis occurring from the ferromagnetic domain effect.

Figure 2d,e,f show the $\gamma$-, $\beta$-, and $\alpha$-dependent AMR of the $y = 1$ superlattice at different $H$ fields at 5 K, respectively. As expected from Figure 2a, b, a larger $H$-field systematically facilitates perpendicular spin ordering, shifting $\gamma_{min}$ and $\beta_{min}$ away from 90° (Figure 2d, e). The consistent $\gamma$- and $\beta$-dependence suggests that both $s$-$d$ scattering and spin Hall contributions are negligible (We provide a detailed discussion on eliminating alternate possibilities of anisotropic magnetic transports in Text S2, Supporting Information); hence, $\gamma_{min}$ and $\beta_{min}$ are predominantly determined by anisotropic spin scattering, as discussed previously. In contrast, $\alpha$-dependent AMR consistently exhibits $\alpha_{min} = 90°$ (Figure 2f), regardless of the $H$-field strength, similar to those in previously reported SRO heterostructures.[29] By taking advantage of the fitting procedure (Figure S7, Supporting Information), we extracted the parameters $p$ and $h$ for different $H$-fields, as shown in Figure 2g. A peculiar nonlinear $H$-field-dependent magnetic anisotropy was clearly visible. The difference between $p$ and $h$ is plotted in Figure 2h, which crosses zero at 4.5 T approximately, suggesting a 45° alignment of the spin ordering away from the $z$-axis. Consistently, at $H$ = ~4.5 T, the difference between MR$_p$ ($H$) and MR$_h$ ($H$) also crosses zero (Figure 2i), and $\gamma_{min}$ indeed reaches 45° (Figure 2j). These results unequivocally indicate a large magnetic anisotropy modulation originating from the competition between the spin scattering along the perpendicular and horizontal directions.

As summarized in **Figure 3**a, b, c, $\gamma_{min}$ changes significantly as a function of the external $H$-field, $T$, and atomically controlled interlayer coupling. Figure 3a schematically shows the modulation of $\gamma_{min}$ from the horizontal to the perpendicular direction with increasing $H$-field and $T$ for the $y = 1$ superlattice. Figure 3b shows that $\gamma_{min}$ of the $y = 1$ superlattice with a large interlayer coupling shifts from 90° to 30° at 5 K. As the interlayer coupling strength decreased, the $y = 2$ superlattice exhibited a smaller $H$-field dependence of $\gamma_{min}$ from 90° to 72° (top panels of Figure S8, Supporting Information), whereas the $y = 6$ superlattice maintained $\gamma_{min} = 90°$ irrespective of the $H$-field strength (bottom panels of Figure S8). With decreasing $T$ (< 20 K), $\gamma_{min}$ of the $y = 1$ superlattice with 9 T further rotated from 0° to 30°, whereas those of the $y = 2$ and 6 superlattices remained at 72° and 90°, respectively (Figure 3c). The AMR of the $y = 6$ superlattice in the absence of interlayer coupling provides experimental evidence that the 2D Mott insulating phase of the SRO monolayer inherently possesses a horizontal AMR with $\gamma_{min} = 90°$.

We investigated magnetic anisotropy energy (MAE, i.e., energy difference between the horizontal and perpendicular spin-ordered states, $E_{//} - E_{\perp}$) to theoretically reveal the possible origin of magnetic anisotropy modulation in 2D SRO crystal, as shown in Figure 3d, e, and Figure S10. Figure 3d shows that all of [1|y] superlattices with y = 1, 2, and 6 exhibit a negative MAE. This indicates that the spin ordering of SRO monolayers within superlattices preferentially aligns along the horizontal direction as the ground state, in contrast to thicker films. The [1|6] superlattice, which approaches the 2D limit, exhibits the largest negative MAE value corroborating the experimentally observed AMR with $\gamma_{min}$ = 90°. The large negative MAE increases to −7.5 meV/Ru for the y = 2 superlattice and −1.4 meV/Ru for the y = 1 superlattice. In particular, 1.4 meV/Ru is similar to the thermal energy scale, which is estimated to be ~16 K, consistent with the T at which the horizontal AMR appears. In addition, by adopting the H-field dependent spin Hamiltonian for layered magnetic materials,[29] H-field induced MAE is expected to be ~1 meV/Ru for 2 $\mu_B$/Ru of M at 9 T, which is comparable to the calculated MAE for the y = 1 superlattice.

According to the second-order perturbation theory, the MAE is determined by the spin-orbit interaction matrix elements as follows.[31]

$$\text{MAE} = E_{//} - E_{\perp} = \xi^2 \sum_{u.o.\sigma,\sigma'} (-1)^{1-\delta_{\sigma\sigma'}} \left[ \frac{|\langle o^{\sigma'}|L_z|u^{\sigma}\rangle|^2 - |\langle o^{\sigma'}|L_x|u^{\sigma}\rangle|^2}{E_u^{\sigma} - E_o^{\sigma'}} \right]$$

where $\xi$ is the spin-orbit coupling constant, and $L_x$ and $L_z$ are the angular momentum operators along the x (horizontal) and z (perpendicular) directions, respectively. The $|o\rangle$ and $|u\rangle$ are the occupied and unoccupied orbital states, respectively; $E_o$ and $E_u$ are the eigenenergies of the $|o\rangle$ and $|u\rangle$, respectively; and $\sigma$ indicates the spin state. Here, the MAE can be decomposed into matrix elements indexed as (o, u), and we select five dominant matrix elements, as shown in Figure 3e (all matrix elements are given in Figure S9, Supporting Information). In particular, as y increases (corresponding to a decrease in interlayer coupling strength), the MAE of the ($d_{xy}$, $d_{yz}$) and ($d_{yz}$, $d_{zx}$) elements within the $t_{2g}$ orbitals decreases systematically, as shown in Figure 3e and Figure S10. The theoretical analyses strongly support that interlayer coupling between SRO monolayers effectively changes the orbital hybridization and resultant MAE via atomically controlled y within the superlattices. More specifically, the strong interlayer coupling between the SRO monolayers in the y = 1 superlattice leads to a nearly degenerate MAE, providing an effective engineering of the magnetic anisotropy and AMR via both the external H-field and T.

## 3. Conclusion

In summary, we demonstrated an atomically controlled dimensional Mott transition and its large AMR modulation in 2D SRO crystals. We employed atomic-scale interlayer coupling control to systematically induced the digitized dimensional Mott transition of monolayer SRO within superlattices. The competition between the antiferromagnetic nature of 2D SRO and interlayer coupling-induced ferromagnetic order led to the huge modulation of AMR via interlayer coupling, $T$, and an external $H$-field. Theoretical calculations confirmed that the magnetic anisotropy of 2D SRO monolayers could be modulated significantly by introducing interlayer coupling. Our observations provide insight into the dimensional Mott transition of correlated systems, which sheds light on the development of future spin-induced Mottronics via magnetic anisotropy control.

## 4. Methods

*DFT calculations*: First-principles DFT calculations were performed using the generalized gradient approximation of Perdew-Burke-Ernzerhof (GGA-PBE) for exchange-correlation functionals[32] and the projector-augmented wave (PAW) method with a plane-wave basis,[33] as implemented in the Vienna ab initio simulation package (VASP) code.[34] Plane waves were included up to a kinetic energy cutoff of 500 eV. For Brillouin-zone integration, Γ-centered 4 × 4 × 2 k-point grid meshes were used for all the calculations. For [1|y] superlattices, the in-plane lattice constant was fixed to $\sqrt{2} \times \sqrt{2} a_{STO}$ ($a_{STO}$ = 3.943 Å), obtained using the PBE potential. The calculations were converged in energy to $10^{-6}$ eV/cell, and the structures were allowed to relax until the forces were less than $10^{-2}$ eV/Å. Based on previous experiments of structural characterization using scanning transmission electron microscopy and X-ray diffraction,[16, 19, 21, 35-39] we employed a tetragonal SRO layer structure with suppressed octahedral distortion in [1|y] superlattices. An on-site Coulomb repulsion $U$ of 2.6 eV was applied to the Ru-$d$ orbital. MAE was estimated from the total energy difference between the horizontally and perpendicularly oriented spin systems.

*Atomically designed synthetic SRO crystal*: To realize a 2D SRO system with modulated interlayer coupling, we grew the [x|y] superlattice (x- and y-unit cells of the SRO and STO layers with 10 repetitions) on (001)-oriented single-crystal STO substrates using pulsed laser epitaxy.[16, 19, 21, 35-39] Note that a single perovskite unit cell is 0.4 nm approximately. High-resolution X-ray diffraction (XRD) $\theta$-$2\theta$ scans show clear Bragg peaks of the [1|y] superlattice with different y, corresponding to the atomically designed periodicity of supercell structures (Figure S1 in the Supporting Information). The atomic force microscope image of the [1|1] superlattice typically shows a step-terrace structure, which indicates the atomically sharp surface of the superlattice, as shown in the inset of Figure S1. To confirm the y-dependence of the octahedral distortion of the superlattices, we measured half-integer Bragg peaks using synchrotron XRD at the 5D beamline of the Pohang Light Source at room temperature (Figure S2, Supporting Information). Note that we could not find any structural variation of the [1|y] superlattices depending on y, which is distinct from the previously reported SRO/STO superlattices with 50 repetitions.[40]

*Magnetization and electrical transport measurements*: Magnetization measurements of the SRO/STO superlattices were performed using a magnetic property measurement system (MPMS, Quantum Design). $M(T)$ was measured using field-cooled methods along the perpendicular directions. $M(H)$ was recorded with horizontal and perpendicular $H$-fields at 5 K. The electrical resistance and MR were measured using a Physical Property Measurement System (Quantum Design, PPMS) with a built-in horizontal rotator. Contacts of Au wires were attached to the edges of the sample (0.5 × 0.5 cm$^2$ of typical size) with indium electrodes for the Van der Pauw configurations. Polar plots of the AMR are shown in Figure S11-S13.

*X-ray absorption spectroscopy and XMCD*: X-ray absorption spectroscopy (XAS) and XMCD at the Ru-$L_{2,3}$ edge of the [1|1] superlattice were measured at the 6A beamline of the Pohang Light Source using on-the-fly methods. XAS spectra were measured at room $T$ and XMCD spectra were recorded at 3 K with 7 T of $H$-field.

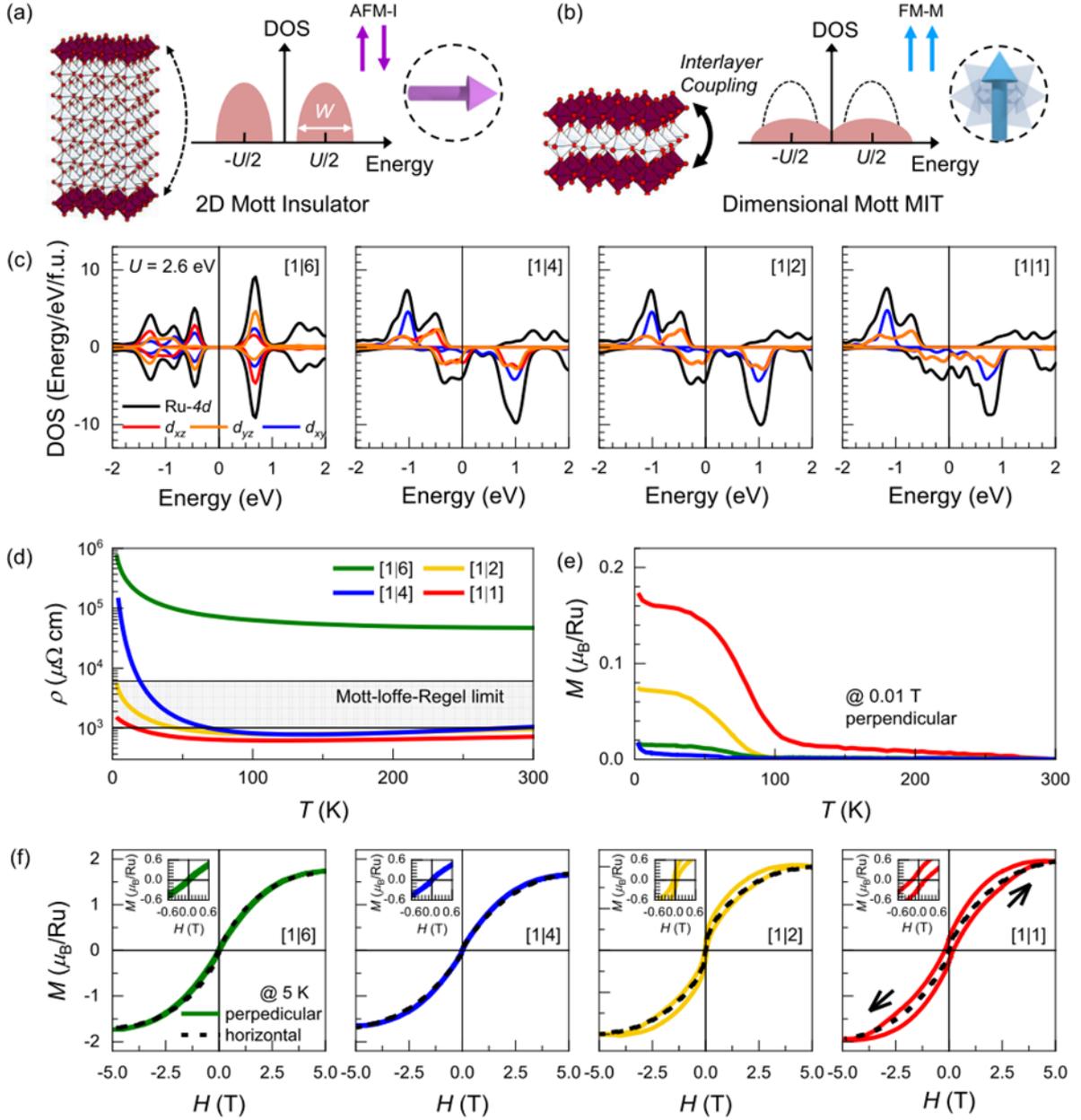

**Figure 1. Dimensional Mott MIT with magnetic phase transition in atomically controlled monolayer SRO superlattices.** a,b) Schematic illustration of (a) 2D Mott insulator and (b) dimensional Mott MIT in atomically controlled monolayer SRO superlattices. c) DFT calculations show interlayer coupling induced dimensional Mott MIT with magnetic phase transition. The positive and negative signs of density of state (DOS) indicate the spin up and down. d,e) $T$-dependent (d) resistivity and (e) $M$ for [1|y] superlattices. The Mott-Ioffe-Regel limit is determined by $k_F\lambda \sim 1 - 2\pi$, where $k_F$ and $\lambda$ are Fermi wave vector and mean free path of the electrons in the SRO layer, respectively. f) perpendicular and horizontal $M(H)$ curves for [1|y] superlattices. The arrows indicate the scan direction of the $H$-field. The inset shows the extended $M(H)$ curves near the 0 T.

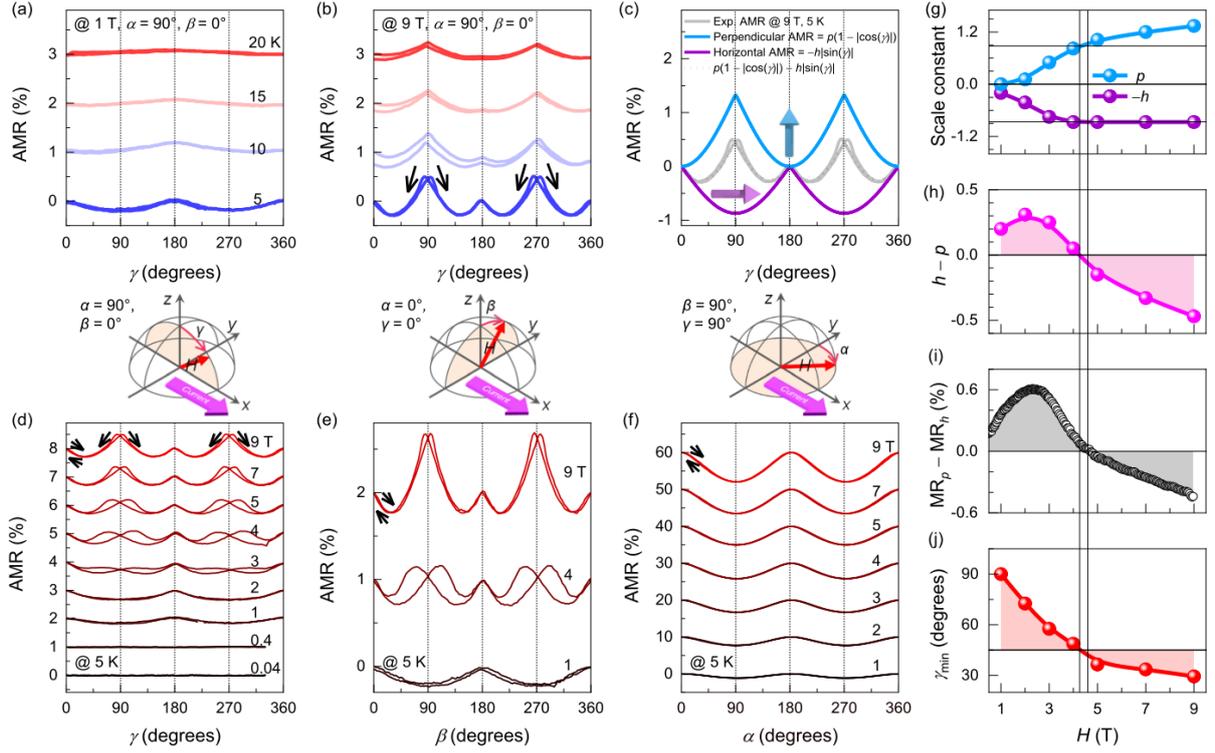

**Figure 2.** *H*-field-induced of AMR modulation in atomically controlled monolayer SRO superlattices. a,b) *T*-dependent AMR (*γ*) with (a) 1 T and (b) 9 T of *H*-field below 20 K. c) Simple consideration of the two different AMR originating from anisotropic horizontal (perpendicular) spin ordering schematically shown in (c). The second row shows the schematic experimental configurations of AMR with three different *H*-field directions. The violet arrows indicate the direction of the current. d-f) *H*-field dependent AMR results of *y* = 1 superlattice along the (d) *α*, (e) *β*, and (f) *γ* direction. The arrows indicate the scan direction of the *H*-field. We include the 1 (10) % offset for AMR (*γ*) and AMR (*β*) (AMR (*α*)). g-j, Summary of the (g) scale constant, (h) $h - p$, (i) $MR_p - MR_h$, and (j) $γ_{min}$ as a function of *H*-field at 5 K for *y* = 1 superlattices.

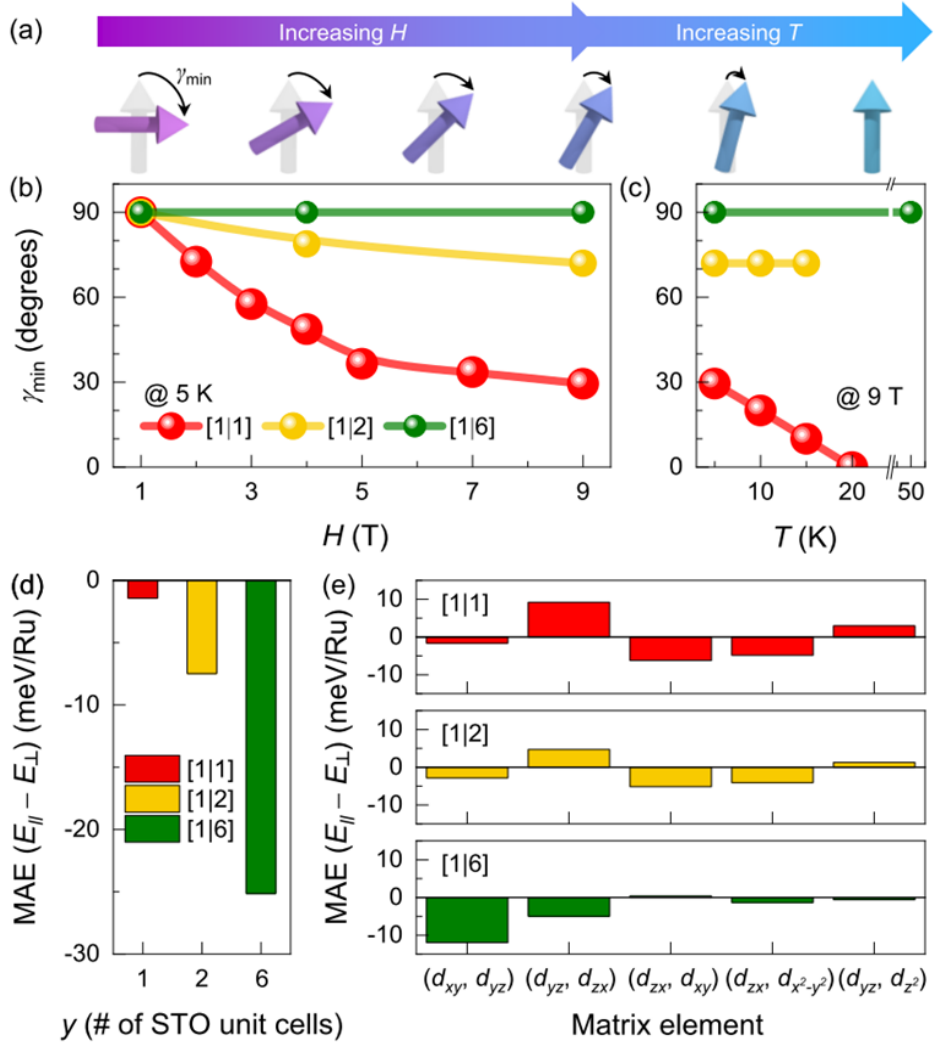

**Figure 3. Summary of $\gamma_{min}$ modulation as functions of external *H*-field, *T*, and atomically controlled interlayer coupling strength and its theoretical understanding.** a-c) (a) Schematic illustration of *H*- and *T*-dependent $\gamma_{min}$ of for *y* = 1 superlattice. Summary of (b) $\gamma_{min}$ (*H*) at 5 K and (c) $\gamma_{min}$ (*T*) at 9 T for [1|*y*] superlattices. d) The calculated MAE for *y* = 1, 2, and 6 superlattices. e) Decomposition of MAE into dominant five matrix elements.

**Acknowledgements**

This work was supported by the Basic Science Research Programs through the National Research Foundation of Korea (NRF-2021R1A2C2011340, 2022R1C1C2006723, 2018R1A2B6004394, 2022M3H4A1A04085306, 2021R1A2C2010925, and 2020M3H4A2084417). We also thank Core Research Facilities, Pusan National University for MPMS. Seung Gyo Jeong, Jihyun Kim, and Taewon Min contributed equally to this work.

**Author Contributions**

S.G.J. led the research, synthesized and characterized the samples, interpreted results, and wrote the manuscript. T.M. performed theoretical calculations under the supervision of J.L. S.G.J. and J.K. performed electrical transport measurements under the supervision of T.P. S.S. and S.P. conducted magnetization measurements. S.G.J., J.O., and W.-s.N. conducted X-ray absorption spectroscopy. J.L. and W.S.C. supervised the research, interpreted the results, and wrote the manuscript. All authors contributed to the preparation of the manuscript and discussion of the results.

# Supporting Information

**Exotic magnetic anisotropy near digitized dimensional Mott boundary**

Seung Gyo Jeong, Jihyun Kim, Taewon Min, Sehwan Song, Jin Young Oh, Woo-suk Noh, Sungkyun Park, Tuson Park, Jong Mok Ok, Jaekwang Lee*, and Woo Seok Choi*

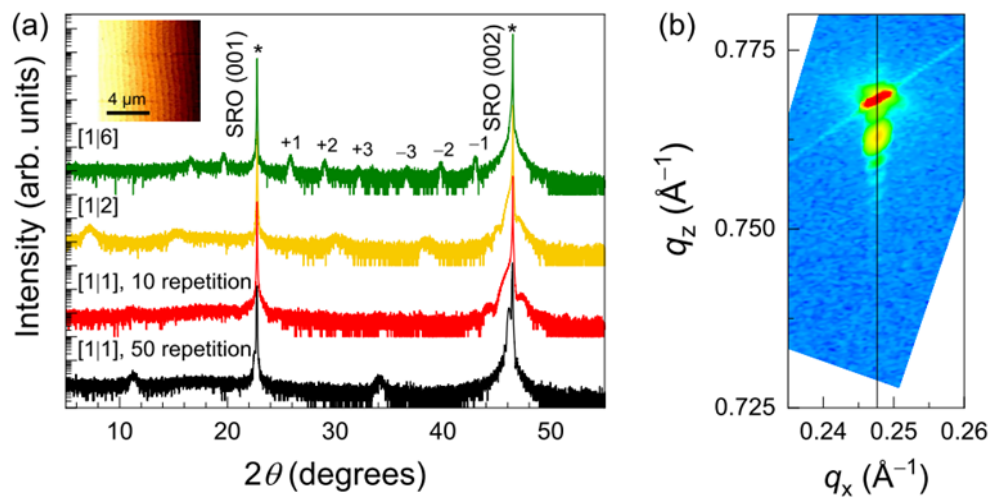

**Figure S1. Structural characterization of atomically designed SRO monolayer superlattice.** a) X-ray diffraction $\theta$-$2\theta$ results of [1|$y$] superlattices with different $y$ show clear Bragg superlattice peaks ($\pm n$) with Pendellösung fringes and their peak position depends on the $y$, representing the atomically well-defined periodicity of superlattices. To confirm the superlattice peak for atomically thin supercells, we synthesized [1|1] superlattice with 50 repetitions (black line). The inset of the atomic force microscopy image shows the typical step-terrace surface of [1|1] superlattice, which represents the atomically flat surfaces of superlattices. b) XRD Reciprocal space map of [1|1] superlattices near the STO (103) plane confirms the fully strained state. The vertical line is a guide to the eye.

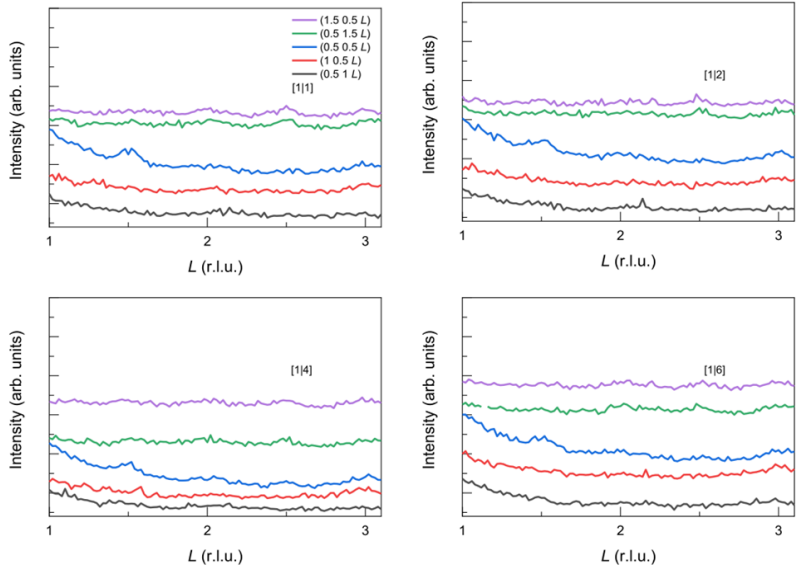

**Figure S2. Half-integer diffraction profile of [1|y] superlattices with different y.**
Synchrotron XRD facility shows small diffraction intensity of half-integer peaks, independence of the $y$. This means that octahedral distortion of [1|y] superlattice is suppressed by epitaxial strain. Thus, we exclude the $y$-dependent structural modulation scenario to possibly change the magnetic anisotropy of SRO monolayers within the superlattice.

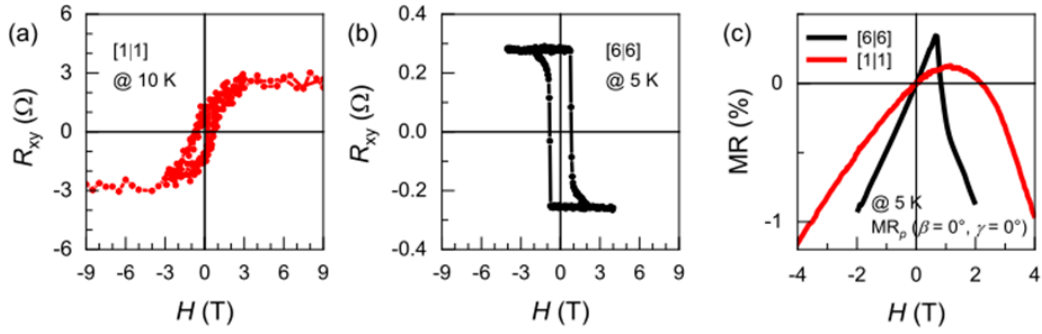

**Figure S3. Ferromagnetic magnetotransport behavior of [1|1] superlattice.** a, b) $H$-field-dependent Hall resistance ($R_{xy}$) (a) [1|1] and (b) [6|6] superlattice shows the typical ferromagnetic hysteresis. The slope of the coercive field of [1|1] superlattice is much broader than that of [6|6] superlattice. c) $MR_p$ as a function of $H$-field ($\beta = 0°$ and $\gamma = 0°$) shows the broader peak curvature of [1|1] superlattice compared to that of [6|6] superlattice.

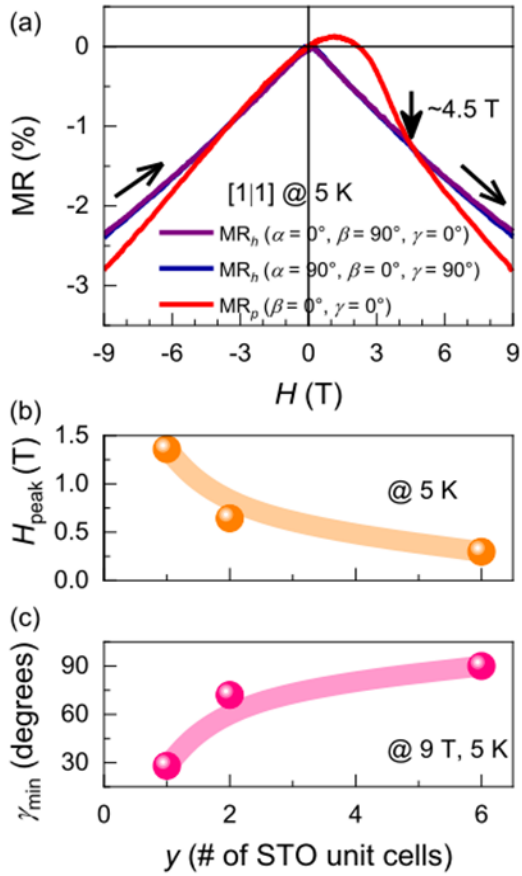

**Figure S4. *H*-field dependent MR$_h$ and MR$_p$ for [1|1] superlattice.** a) *H*-field dependent MR$_h$ and MR$_p$ for [1|1] superlattice cross at ~4.5 T. The arrow indicates the sweep direction of the *H*-field. b, c) As *y* increases for [1|y] superlattices, (b) $H_{peak}$ decreases whereas (c) $\gamma_{min}$ increases. $H_{peak}$ is defined as *H*-field at maximum MR and $\gamma_{min}$ is the angle of the *H*-field at minimum AMR.

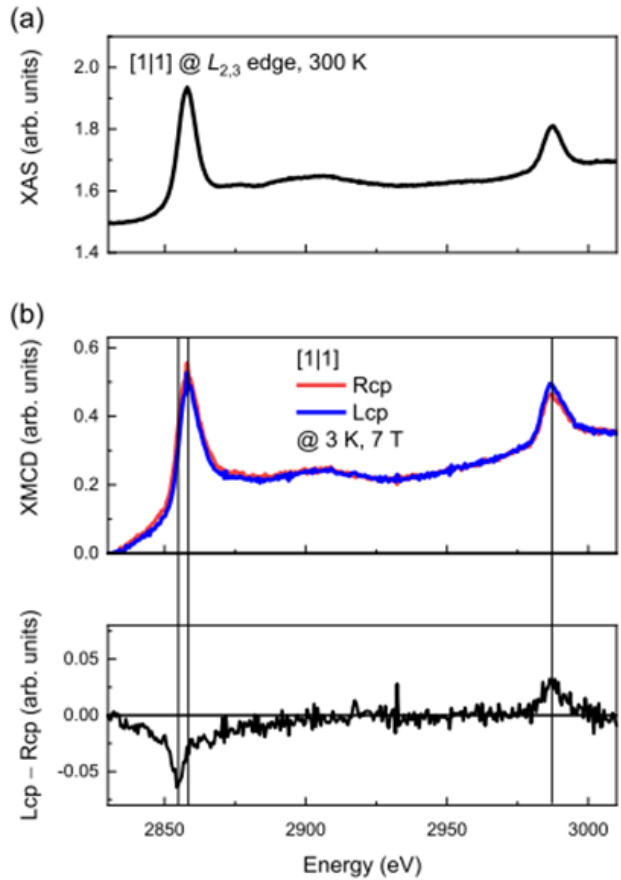

**Figure S5. XAS and XMCD spectra for [1|1] superlattice.** a) XAS spectra for [1|1] superlattice at room-$T$ show the typical electronic structures of Ru $4d$ orbitals states, evidencing the Ru$^{4+}$ oxidation state. B) XMCD spectra at 3 K and 7 T of $H$-field show the finite signal difference between left and right circular polarization (Lcp – Rcp, bottom panel), similar to the previous XMCD study of ferromagnetic SRO thin films.[46] The sample is rotated by 30 (0) degrees from the incident direction of the photon for XMCD (XAS) measurement. External $H$-field is applied along the incident direction of the photon. The XMCD signal confirms that the ferromagnetic behavior is originating from the intrinsic property of SRO and not from possible impurities in the sample.

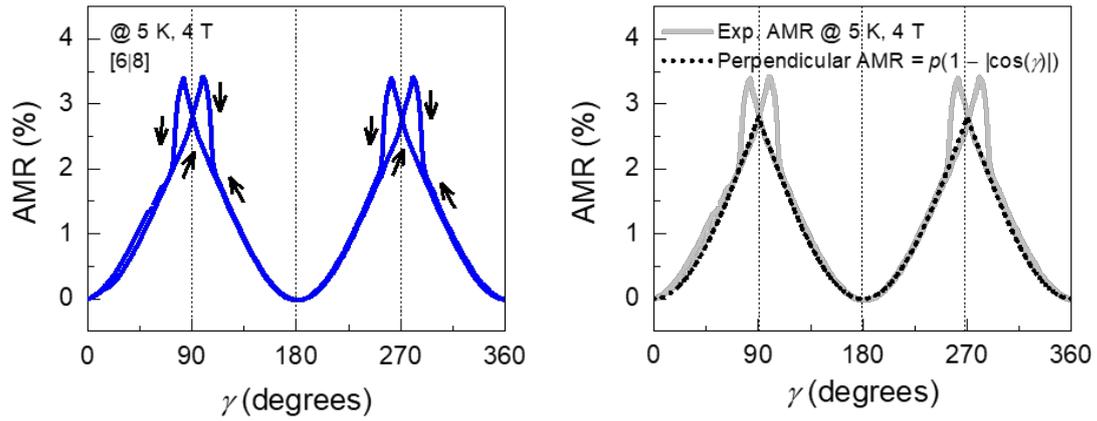

**Figure S6. γ-dependent AMR of [6|8] superlattice.** (Left panel) γ-dependent AMR ($\alpha = 0°$ and $\beta = 0°$) of [6|8] superlattice, where ferromagnetic tetragonal SRO layers, is measured at 4 T and 5 K along the $\beta = 0°$ and $\gamma = 0°$). (Right panel) Perpendicular AMR model ($p(|1 - \cos(\gamma)|)$) accurately describes the 3D tetragonal SRO, except for the ferromagnetic hysteresis.

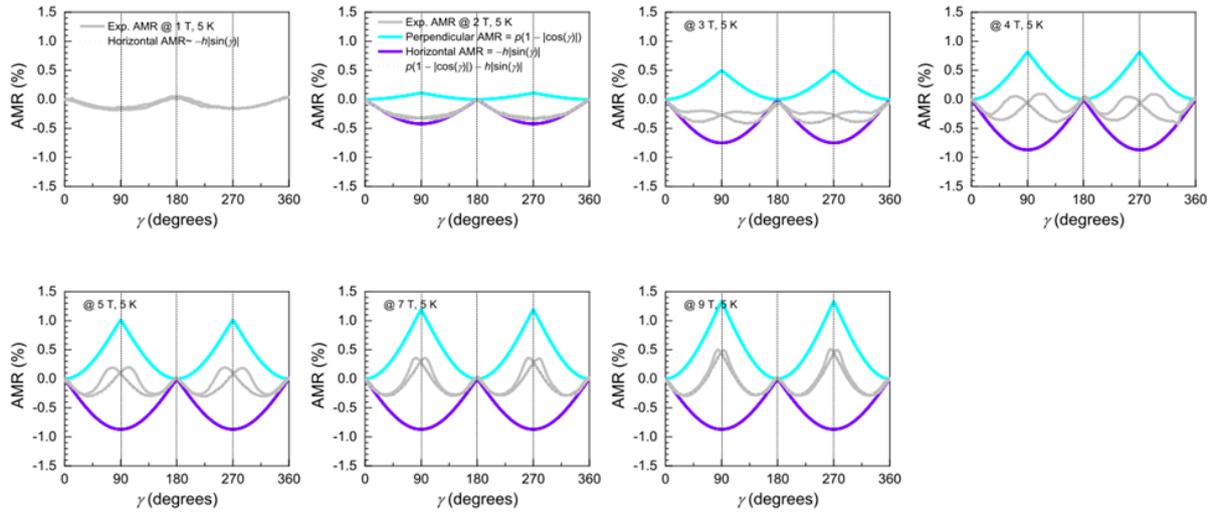

**Figure S7. Fitting results of *H*-dependent AMR (*γ*) with *α* = 0° and *β* = 0° of [1|1] superlattice.** *H*-field dependent AMR (*γ*) of [1|1] superlattice can be explained by the combination of horizontal and perpendicular AMR models (see Text S1 in the Supporting Information for more detail).

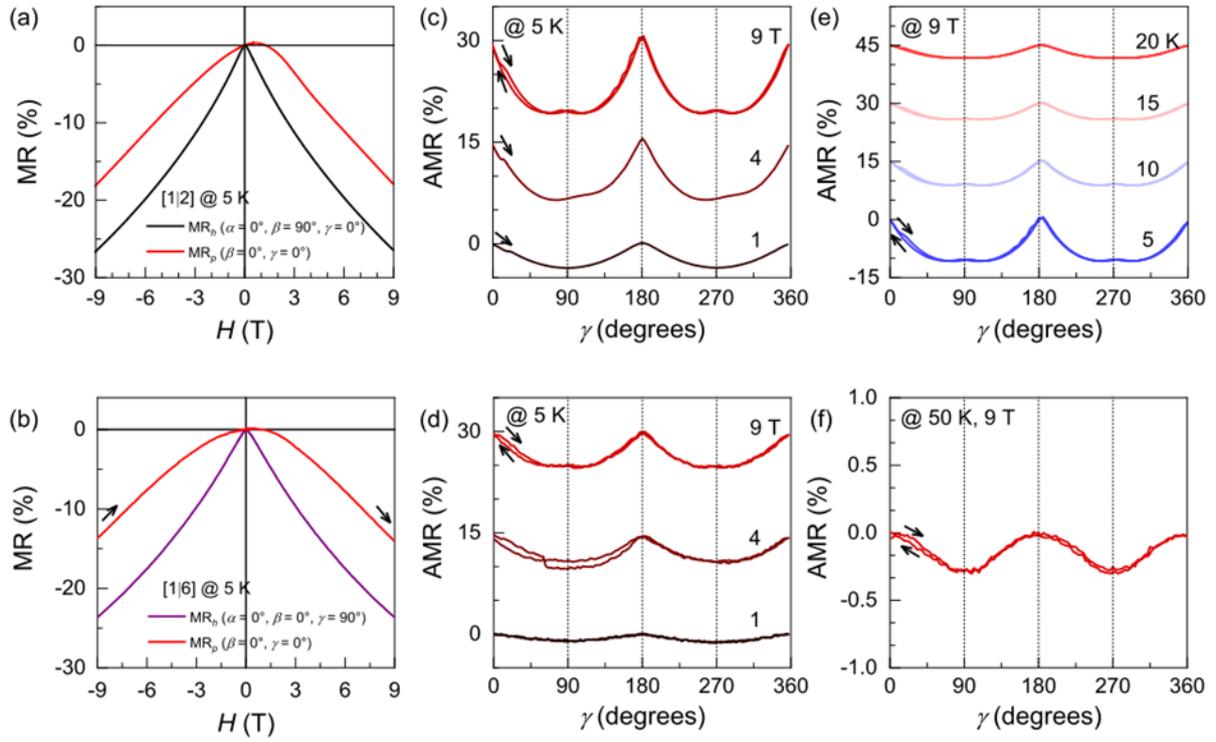

**Figure S8. MR and AMR curves for [1|2] and [1|6] superlattice.** a-d) *H*-field dependent MR$_h$ and MR$_p$ for (a) [1|2] and (b) [1|6] superlattices at 5 K. *H*-field dependent AMR (γ) (α = 0° and β = 0°) for (c) [1|2] and (d) [1|6] superlattices at 5 K. e) *T*-dependent AMR (γ) for [1|2] superlattice with 9 T of *H*-field. f, AMR (γ) for [1|6] superlattice at 50 K and 9 T.

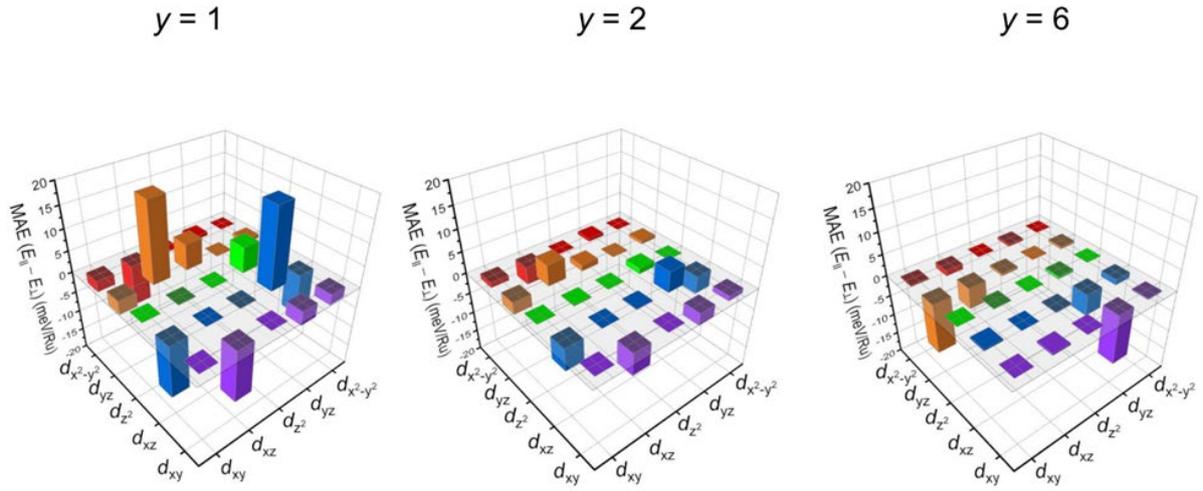

**Figure S9. Extended matrix elements of MAE calculations for [1|*y*] superlattices with different *y*.** ($o$, $u$) element denotes the MAE component between $|o\rangle$ and $|u\rangle$ orbital states (see the discussion in the Main Text for more detail).

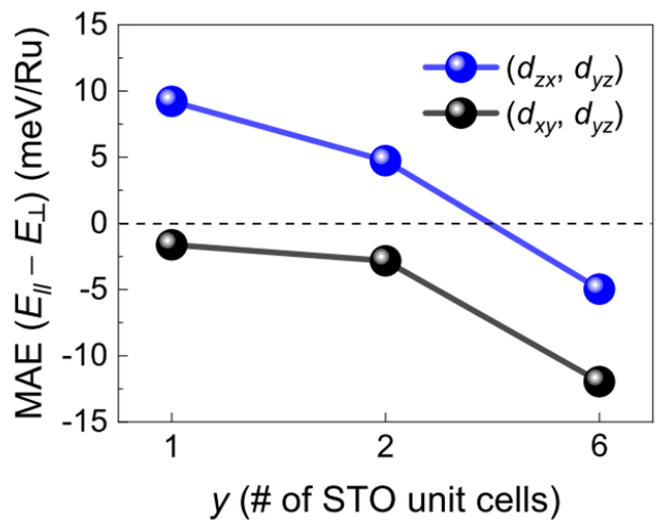

**Figure S10. Dominant two matrix elements (($d_{yz}$, $d_{zx}$) and ($d_{xy}$, $d_{yz}$)) of MAE for [1|y] superlattices with different y.**

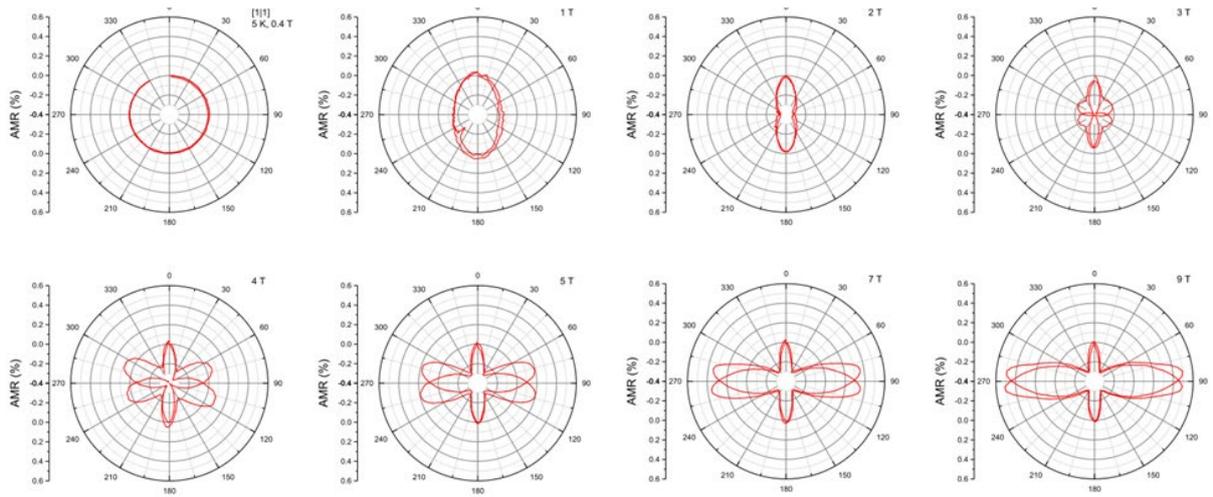

**Figure S11. Polar plots of *H*-field dependent AMR (*γ*) at 5 K for [1|1] superlattice.**

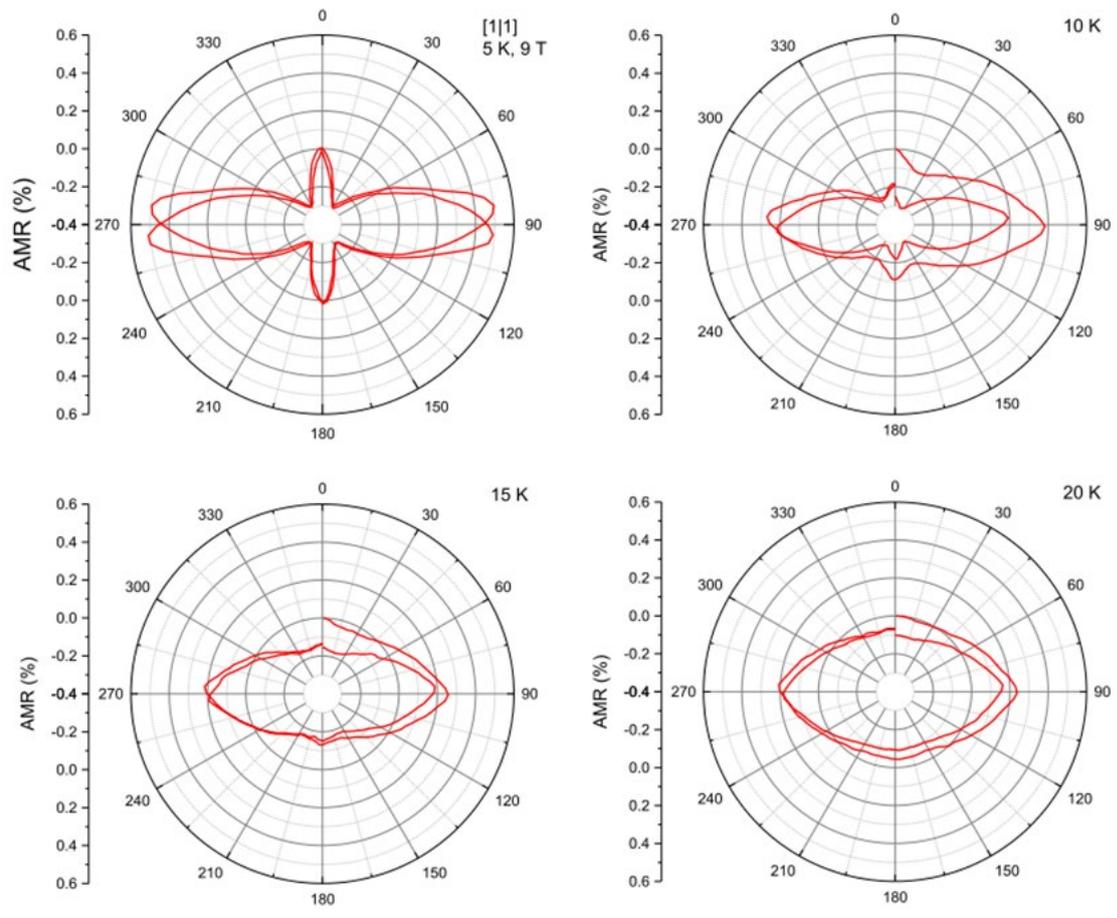

**Figure S12. Polar plots of *T*-dependent AMR (*γ*) with 9 T for [1|1] superlattice.**

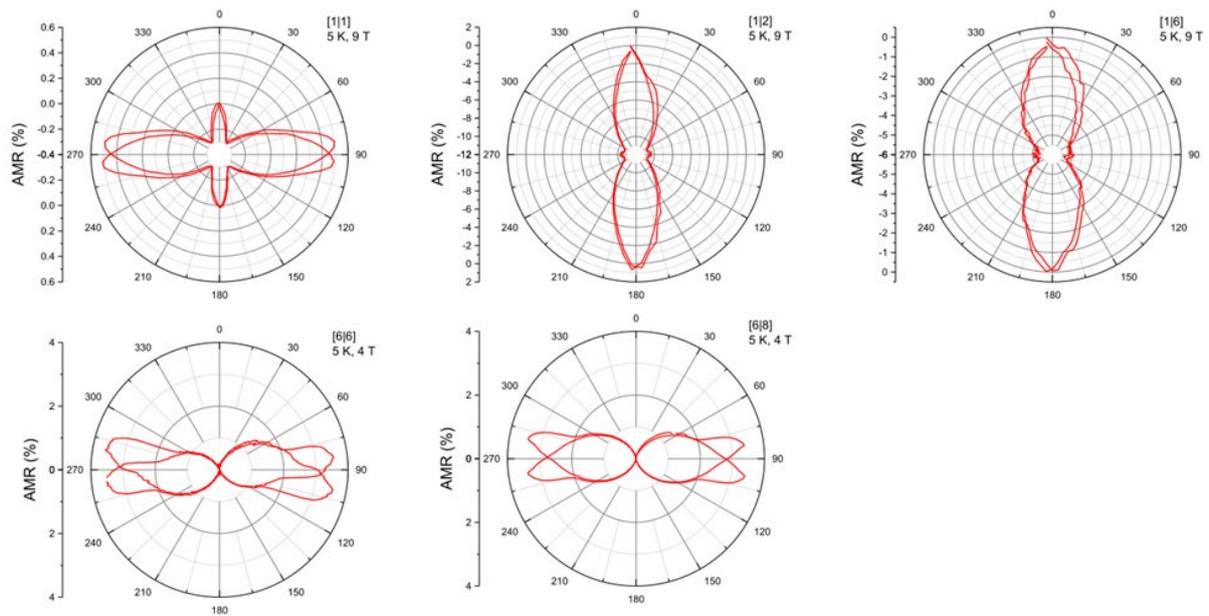

**Figure S13. Polar plots of *y*-dependent AMR (*γ*) at 5 K and 9 T for [1|*y*] superlattice with different *y*.** The bottom panels show the AMR (*γ*) at 5 K and 4 T for tetragonal [6|6] and [6|8] superlattices for comparisons.

**Text S1. Model consideration of AMR.**

To understand the complex AMR evolution depending on the external $H$-field and $T$, we decomposed the AMR in terms of perpendicular ($p(|1 - \cos(\gamma)|)$) and horizontal components ($-h|\sin(\gamma)|$), as shown in Figure 3c and Figure S7, S8. First, we note that perpendicular AMR has been reported in various conventional 3D SRO systems.[19, 27-29] The tetragonal SRO shows a clear perpendicular AMR with $\gamma_{min} = 0°$, which is closely related to the perpendicular easy axis, whereas orthorhombic SRO shows a similar perpendicular AMR with a small offset ($\gamma_{min} \neq 0°$) originating from the distorted $RuO_6$ octahedra. Figure S6 shows that the AMR of tetragonal SRO/STO superlattices with thick SRO layer is well described in terms of $p(|1 - \cos(\gamma)|)$ only, except for the ferromagnetic hysteresis near ~90°. Second, the horizontal AMR component can be explained by $-h|\sin(\gamma)|$, because the curvature of the peak at 0° is much sharper than that of the valley at 90° (Figure 3c). The top left panel of Figure S7 shows that the AMR at 1 T and 5 K for [1|1] superlattice was well fitted by the horizontal AMR component only. With an increasing $H$-field, the combination of perpendicular and horizontal AMR components should be considered to explain the AMR above 2 T at 5 K. We also note that the AMR of the [1|6] superlattice at 5 K can be understood by horizontal AMR component, regardless of the $H$-field strength, similar to the AMR of low-dimensional systems (Figure S8).[41, 42] This analysis validates the existence of anisotropic spin scattering in the monolayer SRO superlattices. We note that the $H$-field dependent AMR behavior for [1|1] superlattices cannot be understood by the conventional Morin transition (spin reorientation of antiferromagnetic order) in hematite[42] because the AMR of the [1|1] superlattice exhibits unprecedented $H$-field dependent rotation.

**Text S2. Possible microscopic mechanisms of AMR modulation.**

We note that spin ordering alone may not fully explain the MR in several single-SRO thin film cases.[44] For example, MR does not saturate at a high $H$-field, and MR and magnetization measurements may not exhibit the same magnetic anisotropy behavior. Similar discrepancies between the MR and magnetization measurements were also observed in our experiments, suggesting that contributions other than simple spin ordering may exist in our SRO-based superlattices. Nevertheless, our experimental and theoretical results indicate that the most dominant contribution is from the spin ordering. Below, we show that some of the common microscopic magnetotransport mechanisms, such as *s-d* scattering, spin Hall effect, and Lorentz MR, cannot be attributed to the observed behavior.

The longitudinal magnetotransport configuration could strengthen the *s-d* scattering, whereas the transverse configuration might induce the spin Hall effect. Hence, these two effects are mutually exclusive. The AMR of the *y* = 1 superlattice exhibited similar *γ*- and *β*-dependence (Figure 2e, f), suggesting little difference in the overall MR characteristics between the longitudinal and transverse configurations. This implies that both the *s-d* scattering and spin Hall contributions are small. Furthermore, the absence of $H^2$-dependence of the MR with three different directions of the *H*-field supports a small contribution from the Lorentz force (Figure S4).[44]